\documentclass[reprint,amsmath,amssymb,aps]{revtex4-1}
\usepackage[colorlinks,citecolor=blue,linkcolor=blue,anchorcolor=blue,filecolor=blue,
urlcolor=blue]{hyperref}
\usepackage{graphicx}
\usepackage{ulem}
\usepackage{dcolumn}
\usepackage{bm}
\usepackage{todonotes}

\begin{document}

\title{Dark matter effects on tidal deformabilities and moment of inertia in a hadronic model with short-range correlations}

\author{O. Louren\c{c}o, C. H. Lenzi, T. Frederico, and M. Dutra}  
\affiliation{Departamento de F\'isica, Instituto Tecnol\'ogico de Aeron\'autica, DCTA, 12228-900, S\~ao Jos\'e dos Campos, SP, Brazil}

\date{\today}

\begin{abstract}
In this work we study the outcomes related to dimensionless tidal deformability $(\Lambda)$ obtained through a relativistic mean-field (RMF) hadronic model including short-range correlations (SRC) and dark matter (DM) content [{\color{blue}Phys. Rev. D {\bf 105}, 023008 (2022)}]. As a dark particle candidate, we use the lightest neutralino interacting with nucleons through the Higgs boson exchange. In particular, we test the model against the constraints regarding the observation of gravitational waves from the binary neutron star merger GW170817 event provided by LIGO and Virgo collaboration (LVC). We show that $\Lambda$ decreases as the dark particle Fermi momentum ($k_F^{\mbox{\tiny DM}}$) increases. This feature favors the \mbox{RMF-SRC-DM} model used here to satisfy the limits of $\Lambda_{1.4}=190^{+390}_{-120}$ ($\Lambda$ of a $1.4M_\odot$ neutron star), and $\tilde{\Lambda}=300^{+420}_{-230}$ given by the LVC. We also show that as $k_F^{\mbox{\tiny DM}}$ increases, $\Lambda_1$ and $\Lambda_2$, namely, tidal deformabilities of the binary system, are also moved to the direction of the GW170817 observational data. Finally, we verify that the inclusion of DM in the system does not destroy the \mbox{$I$-Love} relation (correlation between $\Lambda$ and dimensionless moment of inertia, $\bar{I}$). The observation data for $\bar{I}_\star\equiv\bar{I}(M_\star)=11.10^{+3.68}_{-2.28}$, with $M_\star=1.338M_\odot$, is attained by the \mbox{RMF-SRC-DM} model.
\end{abstract}


\maketitle

\section{Introduction} 
The study of strongly interacting matter at high density regime, i.e., regions whose densities attain some times the nuclear matter saturation density ($\rho_0$), can be performed through the analysis of astrophysical compact objects such as neutron stars (NSs). In the last years, a huge amount of data related to these systems was provided by the LIGO/Virgo Collaboration (LVC) since their first detection of gravitational waves (GW), a phenomenon predicted by Albert Einstein in 1916 after the formulation of General Relativity~\cite{einstein1,einstein2}. LVC published in Ref.~\cite{bholes1} their results regarding the GW produced by two colliding black holes, detected in 2015 in an event called GW150914. In 2017 the event GW170817, also detected by LVC~\cite{ligo17}, was confirmed to have produced GW from the merger of two NSs in a binary system. This former event gave rise to constraints on the tidal deformabilities of each companion star, namely, an effect analogous to the tides observed in our planet due to presence of the Moon. 

Neutron stars can also be used as source of study of dark matter (DM)~\cite{dmrev,zwicky,oort,lensing}. Despite the interaction between dark particles and luminous matter is extremely weak (otherwise DM would be easily detected), the gravitational force can bound this exotic matter to the ordinary one present in the massive astrophysical systems. In that direction, many investigations were performed in which DM is coupled to hadronic relativistic mean field (RMF) models, see Refs.~\cite{rmfdm1,rmfdm2,rmfdm3,rmfdm4,rmfdm5,rmfdm6,rmfdm7,rmfdm8,rmfdm9,rmfdm10,rmfdm11,rmfdm12} for instance. In the most of these studies. the lightest neutralino, belonging to a class of weakly interacting massive particles (WIMP)~\cite{cand1,cand2}, is used as dark particle candidate, but there are other ones, namely, gravitinos, axinos, axions, sterile neutrinos, WIMPzillas, supersymmetric Q-balls, and mirror matter~\cite{cand1,cand2}. 

In Ref.~\cite{dmnosso} the lightest neutralino was implemented as the dark particle in a RMF model with short-range correlations (SRC)~\cite{nature,hen2017,Duer2019,rev3,cai,baoanlicai} included. These kind of correlations are observed in electron-induced quasielastic proton knockout experiments, in which nonindependent nucleons are verified to correlate each other in pairs with high relative momentum. Probes of this phenomenon were performed in experiments at the Thomas Jefferson National Accelerator Facility (JLab), where it was found that the most of the emerged pairs are deuteron-like~\cite{orhen}, around $90\%$ in measurements of the \mbox{$^{12}$C nucleus}~\cite{subedi2008}, for instance. Based on this \mbox{RMF-SRC} hadronic model, it was shown in Ref.~\cite{dmnosso} that it is possible to describe NSs with DM content presenting masses in the limits given in Ref.~\cite{cromartie}, namely, $M=2.14^{+0.10}_{-0.09}M_{\odot}$ ($68.3\%$ credible level), and simultaneously in agreement with the recent observational data provided by the NASA’s Neutron star Interior Composition Explorer (NICER) mission that, provided constraints on the mass-radius profiles~\cite{nicer1,nicer2,nicer3,nicer4}. The ``best'' parametrizations of this \mbox{RMF-SRC-DM} model were constructed by taking into account the variation in the symmetry energy slope in a range compatible with the results reported by the updated lead radius experiment~(PREX-2)~\cite{piekaprex2,prex2}, and also overlapping with the boundaries obtained from the analysis of charged pions spectra~\cite{pions}. The results provided in Ref.~\cite{dmnosso} are also compatible with more recent data given in Ref.~\cite{cromartie-apj} regarding the most massive neutron star known, namely, $M=2.08^{+0.07}_{-0.07}M_{\odot}$ ($68.3\%$ credibility).

In this work, we investigate whether it is also possible to describe the constraints related to the  dimensionless tidal deformabilities regarding the GW170817 event by using the \mbox{RMF-SRC-DM} model of Ref.~\cite{dmnosso}. In particular, we verify that the system with DM content is able to reproduce the limits of $\Lambda_{1.4}=190^{+390}_{-120}$~\cite{ligo18} (dimensionless tidal deformability of a $ 1.4M_\odot$ NS), the range of $\tilde{\Lambda}=300^{+420}_{-230}$~\cite{ligo19} (quantity related to the dimensionless deformabilities of the binary system stars: $\Lambda_1$ and $\Lambda_2$), and the $\Lambda_1\times\Lambda_2$ regions. Furthermore, we also show that the \mbox{$I$-Love} relation is preserved even in the system with DM. Moreover, we found that the model also satisfies the indirect observational data related to the dimensionless moment of inertia, namely, $\bar{I}(M_\star)=11.10^{+3.68}_{-2.28}$, with $M_\star=1.338M_\odot$. Regarding this last quantity, we remark that the obtained range is not coming from a direct measured observable. It is a derived quantity under certain assumptions, as we make clear later on. We organize all these findings as follows. In Sec.~\ref{form}, we address the main equations regarding the \mbox{RMF-SRC-DM} model. The predictions of the model concerning the GW170817 constraints on the tidal deformabilities and moment of inertia are given in Sec.~\ref{stellar}. Our summary and concluding remarks are presented in Sec.~\ref{summ}.

\section{Hadronic model with SRC and  DM } 
\label{form}

We start by presenting the model that describes the hadronic matter considered here, defined by its  Lagrangian density. It reads
\begin{align}
&\mathcal{L}_{\mbox{\tiny HAD}} = \overline{\psi}(i\gamma^\mu\partial_\mu - M_{\mbox{\tiny nuc}})\psi 
+ g_\sigma\sigma\overline{\psi}\psi 
- g_\omega\overline{\psi}\gamma^\mu\omega_\mu\psi
\nonumber \\ 
&- \frac{g_\rho}{2}\overline{\psi}\gamma^\mu\vec{\rho}_\mu\vec{\tau}\psi
+\frac{1}{2}(\partial^\mu \sigma \partial_\mu \sigma - m^2_\sigma\sigma^2)
- \frac{A}{3}\sigma^3 - \frac{B}{4}\sigma^4 
\nonumber\\
&-\frac{1}{4}F^{\mu\nu}F_{\mu\nu} 
+ \frac{1}{2}m^2_\omega\omega_\mu\omega^\mu 
+ \frac{c}{4}(g_\omega^2\omega_\mu\omega^\mu)^2 -\frac{1}{4}\vec{B}^{\mu\nu}\vec{B}_{\mu\nu} 
\nonumber \\
&+ \frac{1}{2}\alpha'_3g_\omega^2 g_\rho^2\omega_\mu\omega^\mu
\vec{\rho}_\mu\vec{\rho}^\mu + \frac{1}{2}m^2_\rho\vec{\rho}_\mu\vec{\rho}^\mu.
\label{dlag}
\end{align}
In this expression $\psi$ represents the nucleon field, and $\sigma$, $\omega^\mu$, and $\vec{\rho}_\mu$ are the scalar, vector, and isovector-vector fields related to the mesons $\sigma$, $\omega$, and $\rho$, respectively, with tensors $F_{\mu\nu}$ and  $\vec{B}_{\mu\nu}$ given by $F_{\mu\nu}=\partial_\nu\omega_\mu-\partial_\mu\omega_\nu$ and $\vec{B}_{\mu\nu}=\partial_\nu\vec{\rho}_\mu-\partial_\mu\vec{\rho}_\nu$. The mesons masses are $m_\sigma$, $m_\omega$, and $m_\rho$. $M_{\mbox{\tiny nuc}}$ is the nucleon rest mass. Regarding the inclusion of the dark matter content, we proceed as in Ref.~\cite{dmnosso} and consider a dark fermion (mass $M_\chi$, Dirac field $\chi$) interacting with nucleons through the exchange of the Higgs boson (mass $m_h$, scalar field $h$). In this perspective, the Lagrangian density of the total system becomes
\begin{align}
\mathcal{L} &= \overline{\chi}(i\gamma^\mu\partial_\mu - M_\chi)\chi
+ \xi h\overline{\chi}\chi +\frac{1}{2}(\partial^\mu h \partial_\mu h - m^2_h h^2)
\nonumber\\
&+ f\frac{M_{\mbox{\tiny nuc}}}{v}h\overline{\psi}\psi + \mathcal{L}_{\mbox{\tiny HAD}},
\label{dlagtotal}
\end{align}
with $fM_{\mbox{\tiny nuc}}/v$ being the Higgs-nucleon coupling ($v=246$~GeV is the Higgs vacuum expectation value). The constant $\xi$ is the strength of the Higgs-dark particle coupling. 

By using the mean-field approximation to the fields, one has $\sigma\rightarrow \left<\sigma\right>\equiv\sigma$, $\omega_\mu\rightarrow \left<\omega_\mu\right>\equiv\omega_0$, $
\vec{\rho}_\mu\rightarrow \left<\vec{\rho}_\mu\right>\equiv \bar{\rho}_{0(3)}$, $h\rightarrow \left<h\right>\equiv h$, that leads to the following field equations
\begin{align}
m^2_\sigma\,\sigma &= g_\sigma\rho_s - A\sigma^2 - B\sigma^3 
\\
m_\omega^2\,\omega_0 &= g_\omega\rho - Cg_\omega(g_\omega \omega_0)^3 
- \alpha_3'g_\omega^2 g_\rho^2\bar{\rho}_{0(3)}^2\omega_0, 
\\
m_\rho^2\,\bar{\rho}_{0(3)} &= \frac{g_\rho}{2}\rho_3 
-\alpha_3'g_\omega^2 g_\rho^2\bar{\rho}_{0(3)}\omega_0^2, 
\\
[\gamma^\mu (&i\partial_\mu - g_\omega\omega_0 - g_\rho\bar{\rho}_{0(3)}\tau_3/2) - M^*]\psi = 0,
\\
m^2_h\,h &= \xi\rho_s^{\mbox{\tiny DM}} + f\frac{M_{\mbox{\tiny nuc}}}{v}\rho_s
\\
(\gamma^\mu &i\partial_\mu - M_\chi^*)\chi = 0,
\end{align}
with $\tau_3=1$ for protons and $\tau_3=-1$ for neutrons. The effective nucleon and dark effective masses are $M^* = M_{\mbox{\tiny nuc}} - g_\sigma\sigma - f\frac{M_{\mbox{\tiny nuc}}}{v}h$ and $M^*_\chi = M_\chi - \xi h$, respectively. Here we use $\xi=0.01$ and $M_{\chi}=200$~GeV (lightest neutralino). Concerning $f$, we use the central value obtained in Ref.~\cite{cline}, namely, $f=0.3$. Such a combination of values gives a spin-independent scattering cross-section around $10^{-47}$~cm$^2$~\cite{rmfdm4} compatible with experimental data from PandaX-II~\cite{pandaxII}, LUX~\cite{lux}, and DarkSide~\cite{darkside} collaborations. Furthermore, the densities are given by $\rho_s=\left<\overline{\psi}\psi\right>={\rho_s}_p+{\rho_s}_n$, $\rho=\left<\overline{\psi}\gamma^0\psi\right> = \rho_p + \rho_n$,
$\rho_3=\left<\overline{\psi}\gamma^0{\tau}_3\psi\right> = \rho_p - \rho_n=(2y_p-1)\rho$, and
$\rho_s^{\mbox{\tiny DM}} = \left<\overline{\chi}\chi\right>$,
where
\begin{eqnarray}
\rho_s^{\mbox{\tiny DM}} &=& 
\frac{\gamma M^*_\chi}{2\pi^2}\int_0^{k_F^{\mbox{\tiny DM}}} \hspace{-0.5cm}\frac{k^2dk}{(k^2+M^{*2}_\chi)^{1/2}}.
\end{eqnarray}
Here $p,n$ defines protons and neutrons, and $\gamma=2$ is the degeneracy factor. The proton fraction is given by $y_p=\rho_p/\rho$, with proton/neutron densities given by $\rho_{p,n}=\gamma{k_F^3}_{p,n}/(6\pi^2)$. ${k_F}_{p,n}$ and $k_F^{\mbox{\tiny DM}}$ are the Fermi momenta related to protons/neutrons, and to the dark particle, respectively.

The thermodynamics of the entire system composed by hadrons and dark matter is determined from the energy density and the pressure, both quantities obtained through the energy-momentum tensor $T^{\mu\nu}$ as $\mathcal{E}=\left<T_{00}\right>$ and $P=\left<T_{ii}\right>/3$. In our case such quantities are given by
\begin{align} 
&\mathcal{E} = \frac{m_{\sigma}^{2} \sigma^{2}}{2} +\frac{A\sigma^{3}}{3} +\frac{B\sigma^{4}}{4} 
-\frac{m_{\omega}^{2} \omega_{0}^{2}}{2} - \frac{Cg_{\omega}^4\omega_{0}^4}{4}
- \frac{m_{\rho}^{2} \bar{\rho}_{0(3)}^{2}}{2} 
\nonumber\\
&+ g_{\omega} \omega_{0} \rho + \frac{g_{\rho}}{2} 
\bar{\rho}_{0(3)} \rho_{3}   -\frac{1}{2} \alpha'_3 g_{\omega}^{2} g_{\rho}^{2} \omega_{0}^{2} 
\bar{\rho}_{0(3)}^{2} + \mathcal{E}_{\mathrm{kin}}^{p} + \mathcal{E}_{\mathrm{kin}}^{n}
\nonumber\\
&+ \frac{m_h^2h^2}{2} + \mathcal{E}_{\mathrm{kin}}^{\mbox{\tiny DM}},
\label{eden}
\end{align}
and
\begin{align}
&P = -\frac{m_{\sigma}^{2} \sigma^{2}}{2} - \frac{A\sigma^{3}}{3} - \frac{B\sigma^{4}}{4} 
+ \frac{m_{\omega}^{2} \omega_{0}^{2}}{2} + \frac{Cg_{\omega}^4\omega_0^4}{4}
\nonumber\\
&+ \frac{m_{\rho}^{2} \bar{\rho}_{0(3)}^{2}}{2} + \frac{1}{2} \alpha'_3 g_{\omega}^{2} 
g_{\rho}^{2} \omega_{0}^{2} \bar{\rho}_{0(3)}^{2} + P_{\mathrm{kin}}^{p} + P_{\mathrm{kin}}^{n}
- \frac{m_h^2h^2}{2} 
\nonumber\\
&+ P_{\mathrm{kin}}^{\mbox{\tiny DM}},
\label{press}
\end{align}
with the kinetic terms of the dark particle written as
\begin{eqnarray}
\mathcal{E}_{\mbox{\tiny kin}}^{\mbox{\tiny DM}} &=& \frac{\gamma}{2\pi^2}\int_0^{k_F^{\mbox{\tiny DM}}}\hspace{-0.3cm}k^2(k^2+M^{*2}_\chi)^{1/2}dk,
\label{ekindm}
\\
P_{\mbox{\tiny kin}}^{\mbox{\tiny DM}} &=& 
\frac{\gamma}{6\pi^2}\int_0^{{k_F^{\mbox{\tiny DM}}}}\hspace{-0.5cm}\frac{k^4dk}{(k^2+M^{*2}_\chi)^{1/2}}.
\label{pkindm}
\end{eqnarray}
In the hadronic side of the system, the implementation of the SRC implies in the replacement of the usual step functions in the kinetic terms by the one including the high momentum tail~\cite{cai,lucas}, namely, $n_{n,p}(k) = \Delta_{n,p}$ for $0<k<k_{F\,{n,p}}$ and $n_{n,p}(k) = C_{n,p}\,(k_{F\,{n,p}}/k)^4$ for $k_{F\,{n,p}}<k<\phi_{n,p} \,k_{F\,{n,p}}$ in which $\Delta_{n,p}=1 - 3C_{n,p}(1-1/\phi_{n,p})$, $C_p=C_0[1 - C_1(1-2y_p)]$, $C_n=C_0[1 + C_1(1-2y_p)]$, $\phi_p=\phi_0[1 - \phi_1(1-2y_p)]$ and $\phi_n=\phi_0[1 + \phi_1(1-2y_p)]$. Here we use $C_0=0.161$, $C_1=-0.25$, $\phi_0 = 2.38$ and $\phi_1=-0.56$~\cite{cai,lucas}. Such change leads to modified expressions to the kinetic terms, namely,
\begin{eqnarray} 
\mathcal{E}_{\text {kin }}^{n,p} &=& \frac{\gamma \Delta_{n,p}}{2\pi^2} \int_0^{{k_{F\,{n,p}}}} 
k^2dk({k^{2}+M^{* 2}})^{1/2}
\nonumber\\
&+& \frac{\gamma C_{n,p}}{2\pi^2} \int_{k_{F\,{n,p}}}^{\phi_{n,p}\, {k_{F\,{n,p}}}} 
\frac{{k_F}_{n,p}^4}{k^2}\, dk({k^{2}+M^{* 2}})^{1/2},
\nonumber \\
P_{\text {kin }}^{n,p} &=&  
\frac{\gamma \Delta_{n,p}}{6\pi^2} \int_0^{k_{F\,{n,p}}}  
\frac{k^4dk}{\left({k^{2}+M^{*2}}\right)^{1/2}} 
\nonumber\\
&+& \frac{\gamma C_{n,p}}{6\pi^2} \int_{k_{F\,{n,p}}}^{\phi_{n,p}\, {k_{F\,{n,p}}}} 
 \frac{{k_F}_{n,p}^4dk}{\left({k^{2}+M^{*2}}\right)^{1/2}},
\end{eqnarray}
and
\begin{align}
&{\rho_s}_{n,p} = 
\frac{\gamma M^*\Delta_{n,p}}{2\pi^2} \int_0^{k_{F\,{n,p}}}  
\frac{k^2dk}{\left({k^{2}+M^{*2}}\right)^{1/2}} 
\nonumber\\
&+ \frac{\gamma M^*C_{n,p}}{2\pi^2} \int_{k_{F\,{n,p}}}^{\phi_{n,p}\, {k_{F\,{n,p}}}} 
\frac{{k_F}_{n,p}^4}{k^2}  \frac{dk}{\left({k^{2}+M^{*2}}\right)^{1/2}}.
\end{align}
This last quantity is the scalar density of protons and neutrons.

\section{Stellar matter: analysis of the GW170817 constraints} 
\label{stellar}

For the description of a NS of mass~$M$ it is needed to solve the widely known TOV equations~\cite{tov39,tov39a} given by  $dp(r)/dr=-[\epsilon(r) + p(r)][m(r) 
+ 4\pi r^3p(r)]/r^2g(r)$ and $dm(r)/dr=4\pi r^2\epsilon(r)$, where $g(r)=1-2m(r)/r$, whose solution 
is constrained to $p(0)=p_c$ (central pressure) and $m(0) = 0$. The condition of $p(R) = 0$ and $m(R)=M$ is satisfied in the star surface, with $R$ defining the NS radius. For the equation of state (EoS) of the matter in the NS core we use the hadronic model with SRC and DM content included. For the NS crust we consider two regions, namely, the outer and the inner crust. For the former we use the EoS proposed by Baym, Pethick and Sutherland (BPS)~\cite{bps} in a density region of $6.3\times10^{-12}\,\mbox{fm}^{-3}\leqslant\rho_{\mbox{\tiny outer}}\leqslant2.5\times10^{-4}\,\mbox{fm}^ {-3}$. For the latter, we follow previous literature~\cite{poly0,poly1,poly2,gogny2,cc2,gogny1,kubis04} and use a polytropic EoS of the form $p(\epsilon)=\mathcal{A}+\mathcal{B}\epsilon^{4/3}$ from $2.5\times10^{-4}\,\mbox{fm}^ {-3}$ to the transition density. The constants $\mathcal{A}$ and $\mathcal{B}$ are found by matching this polytropic formula to the BPS EoS at the interface between the outer and the inner crust, and to the EoS of the homogeneous core at the core-crust transition determined through the thermodynamical method~\cite{gogny1,cc2,kubis04, gonzalez19}. 

In the case of systems composed by binary NS's, the phenomenon of tidal forces originated from the gravitational field takes place, with the consequence of inducing tidal deformabilities in each companion object. The particular deformations due to quadrupole moment produces gravitational waves (GW) whose phase depends on the tidal deformability~\cite{tanj10,read,pozzo}. The first measurement of GW detected from a binary NS's, the called GW170817 event, is due to the LIGO/Virgo Collaboration~\cite{ligo17}. Based on the study related to this new data, the LVC established constraints on the dimensionless tidal deformabilities $\Lambda_1$ and $\Lambda_2$ for each companion star of the binary system, as well as on the tidal deformability related to the star of $M=1.4 M_\odot$ ($\Lambda_{1.4}$). An updated version of the constraints regarding these quantities was published in Refs.~\cite{ligo18,ligo19}. Here we test the capability of the hadronic model with SRC and DM included in satisfying these constraints provided by LVC. In order to do that, we calculate the dimensionless tidal deformability as $\Lambda = 
2k_2/(3C^5)$, with $C=M/R$ (compactness). The second Love number is given by
\begin{eqnarray}
&k_2& = \frac{8C^5}{5}(1-2C)^2[2+2C(y_R-1)-y_R]\nonumber\\
&\times&\Big\{2C [6-3y_R+3C(5y_R-8)] \nonumber\\
&+& 4C^3[13-11y_R+C(3y_R-2) + 2C^2(1+y_R)]\nonumber\\
&+& 3(1-2C)^2[2-y_R+2C(y_R-1)]{\rm ln}(1-2C)\Big\}^{-1},\qquad
\label{k2}
\end{eqnarray}
with $y_R\equiv y(R)$. $y(r)$ is obtained through the solution of $r(dy/dr) + y^2 + yF(r) 
+ r^2Q(r)=0$, solved as part of a coupled system also containing the TOV equations. The quantities $F(r)$ and $Q(r)$ read
\begin{eqnarray}
F(r) &=& \frac{1 - 4\pi r^2[\epsilon(r) - p(r)]}{g(r)} , 
\\
Q(r)&=&\frac{4\pi}{g(r)}\left[5\epsilon(r) + 9p(r) + 
\frac{\epsilon(r)+p(r)}{v_s^2(r)}- \frac{6}{4\pi r^2}\right]
\nonumber\\ 
&-& 4\left[ \frac{m(r)+4\pi r^3 p(r)}{r^2g(r)} \right]^2,
\label{qr}
\end{eqnarray}
where the squared sound velocity is  $v_s^2(r)=\partial p(r)/\partial\varepsilon(r)$. Detailed derivations can be found in Refs.~\cite{tanj10,new,hind08,damour,tayl09}.

The input for the TOV equations coupled to the equation for $y(r)$ is the total equation of state of a system under charge neutrality and $\beta$-equilibrium. In our case, we consider a system composed by protons, neutrons, electrons, muons and dark matter. The total energy density and pressure are given by $\epsilon=\mathcal{E}+\sum_l\epsilon_l$ and $p=P + \sum_lp_l$, with $\mathcal{E}$ and $P$ given in Eqs.~(\ref{eden}) and~(\ref{press}), respectively. The index $l$ refer to the leptons (electons and muons). The equations are solved by taking into account the following conditions: $\mu_n - \mu_p = \mu_e=\mu_\mu$ and $\rho_p - \rho_e = \rho_\mu$, where \mbox{$\rho_l=[(\mu_l^2 - m_l^2)^{3/2}]/(3\pi^2)$} for $l=e, \mu$ (we use $m_e=0$ and  $m_\mu=105.7$~MeV). The chemical potentials of protons, neutrons, electrons and muons are given, respectively, by $\mu_p$, $\mu_n$, $\mu_e$, and $\mu_\mu$. Electron and muon densities are $\rho_e$, and $\rho_\mu$. In the case of the hadronic model with SRC included, $\mu_p$ and $\mu_n$ are given by
\begin{eqnarray} 
&\mu_{p,n}& = 3 C_{p,n} \left[ \mu^{p,n}_{\mathrm{kin}}
- \frac{\left({\phi_{p,n}^2 {k^2_F}_{p,n} + M^{*2}}\right)^{1/2}}{\phi_{p,n}} \right]
\nonumber\\
&+& {4}C_{p,n} {k_F}_{p,n} \ln\left[\frac{\phi_{p,n} {k_F}_{p,n} + 
\left(\phi_{p,n}^2{k_F^2}_{p,n}+M^{*2}\right)^{1/2} }{ {k_F}_{p,n} + \left(  {k^2_F}_{p,n} + M^{*2}\right)^{1/2}}\right] 
\nonumber\\
&+& \Delta_{p,n}\mu^{p,n}_{\mathrm{kin}} + g_{\omega} \omega_{0} \pm \frac{g_\rho}{2}\bar{\rho}_{0_{(3)}},
\end{eqnarray} 
with $\mu^{p,n}_{\mathrm{kin}}=({k^2_F}_{p,n}+M^{*2})^{1/2}$, where we have used the definitions $\mu_{p,n}=\partial\mathcal{E}/\partial\rho_{p,n}$.

As in Ref.~\cite{dmnosso}, we use in the hadronic side of the model the updated version of the parametrization FSU2R~\cite{fsu2r-new}, with the following bulk parameters at the saturation density of symmetric nuclear matter: $\rho_0=0.15$~fm$^{-3}$, $B_0=-16.27$~MeV (binding energy), $M^*_0=556.8$~MeV (effective nucleon mass at $\rho_0$), and $K_0=237.7$~MeV (incompressibility at $\rho_0$). We also use $C=0.004$, $M_{\mbox{\tiny nuc}}=939$~MeV, $m_\sigma=497.479$~MeV, $m_\omega=782.5$~MeV, and $m_\rho=763$~MeV. In Ref.~\cite{dmnosso} the authors have also considered uncertainties in $M_0^*$, $K_0$ and $L_0$ (symmetry energy slope at $\rho_0$). It was verified that changes in $L_0$ produce parametrizations that give mass-radius profiles in agreement with astrophysical observations, such as the boundaries of $M=2.14^{+0.10}_{-0.09}M_{\odot}$~\cite{cromartie}, simultaneously with recent data obtained by the NICER mission~\cite{nicer1,nicer2,nicer3,nicer4}. Here we focus on the variation of this specific isovector quantity. In particular we use~\cite{piekaprex2}
\begin{eqnarray}
L_0=(106\pm37)~\mbox{MeV},
\label{sloperange}
\end{eqnarray}
range compatible with the updated results provided by the \mbox{PREX-2} collaboration concerning neutron skin thickness measurements of $^{208}\rm Pb$~\cite{prex2}, and also overlapping with the limits determined from the analysis of charged pions spectra~\cite{pions}. For each value of $L_0$ chosen in this variation, we fix in $\tilde{J}=25.68$~MeV (FSU2R parametrization) the value of the symmetry energy at $\rho=2\rho_0/3$. This value is consistent with the findings presented in Refs.~\cite{piekaprex2,pieka2001}. By taking this procedure, we impose to the hadronic part of the model the linear correlation between $L_0$ and the symmetry energy at the saturation density, $J$. This is a particular relationship verified in literature, see for instance Refs.~\cite{drischler,baoanli,bianca,wei}.

\begin{figure}[!htb] 
\centering
\vspace{-1cm}
\includegraphics[width=0.54\textwidth]{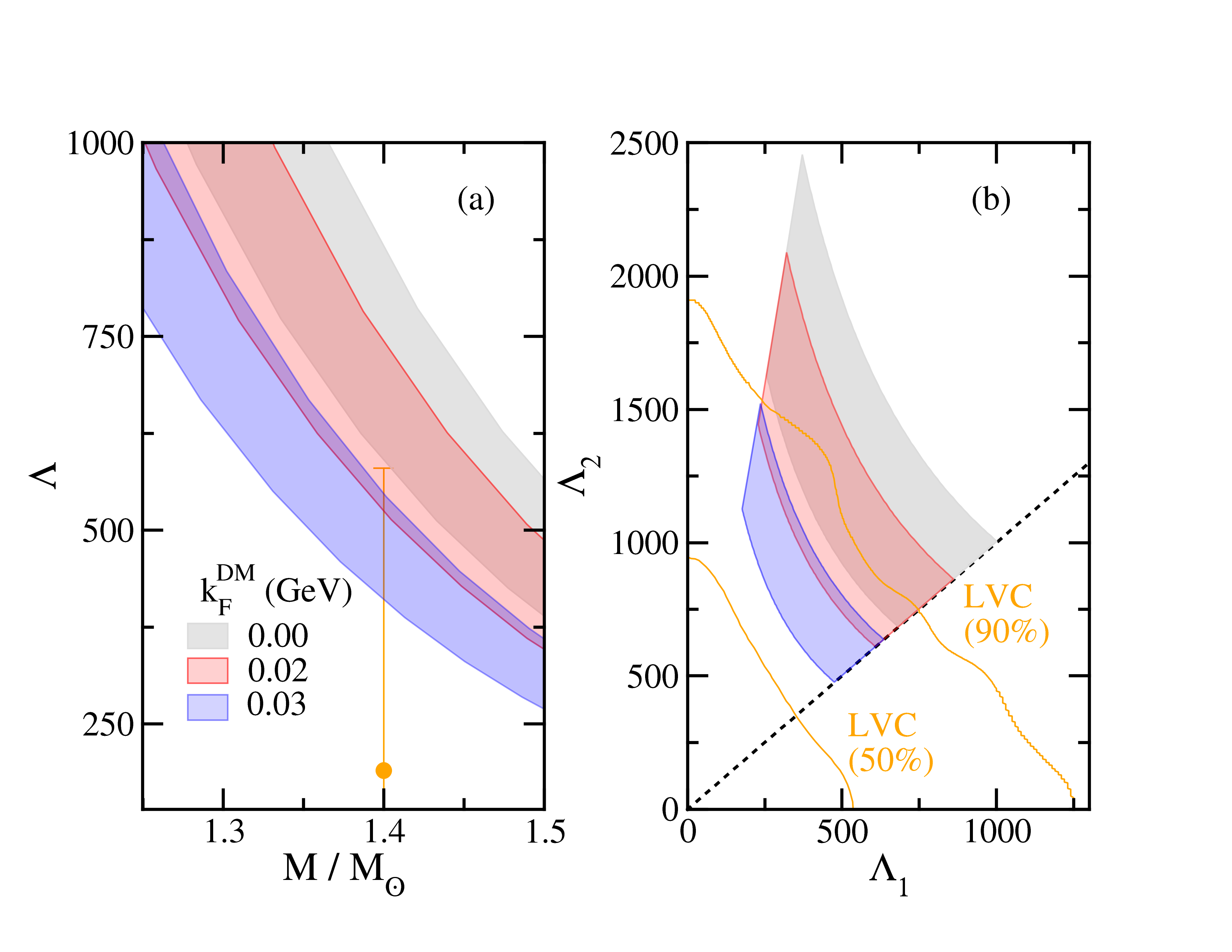}
\vspace{-1cm}
\caption{(a) $\Lambda$ as a function of $M/M_\odot$. Full circle: result of $\Lambda_{1.4}=190^{+390}_{-120}$ obtained in Ref.~\cite{ligo18}. (b) Dimensionless tidal deformabilities for the case of high-mass ($\Lambda_1$) and low-mass ($\Lambda_2$) components of the GW170817 event. Confidence lines, namely, 90\% and 50\%, also taken from Ref.~\cite{ligo18}. For both panels, the dark matter content is characterized by $k_F^{\rm DM}=0$ (no DM: gray bands), $0.02$~GeV (red bands) and $0.03$~GeV (blue bands).} 
\label{def}
\end{figure}

We start by showing in Fig.~\ref{def} the dimensionless tidal deformability generated by the \mbox{RMF-SRC} model with DM included.
In Fig.~\ref{def}{\color{blue}a} we present $\Lambda$ as a function of the NS mass in units of $M_\odot$. Each band represents the set of parametrizations generated by the variation of $L_0$ given in Eq.~(\ref{sloperange}). The content of dark matter is defined by the dark Fermi momentum taken here as~$0$, $0.02$~GeV and $0.03$~GeV. As shown in Ref.~\cite{dmnosso}, $k_F^{\rm DM}=0$ represents the system without dark matter. It is clear that in this case (gray band), the parametrizations obtained by using Eq.~(\ref{sloperange}) do not satisfy the constraint of $\Lambda_{1.4}=190^{+390}_{-120}$~\cite{ligo18}. However, the inclusion of DM favors the system to be compatible with the limit provided by LVC. In particular, for $k_F^{\rm DM}=0.03$~GeV (blue band) it is verified that all parametrizations constructed through Eq.~(\ref{sloperange}) are completely inside the range of $\Lambda_{1.4}$. This value of $k_F^{\rm DM}$ was show in Ref.~\cite{dmnosso} to produce NS's in agreement with the recent observational data regarding the mass-radius diagram. Here we confirm that the system composed by this amount of DM is also consistent with the LVC constraint of $\Lambda_{1.4}$. 

In Fig.~\ref{def}{\color{blue}b} we show the tidal deformabilities $\Lambda_1$ and $\Lambda_2$ of the binary NS's system related to the GW170817 event, with component masses $M_1$, in the range of $1.37\leqslant M_1/M_\odot \leqslant 1.60$~\cite{ligo17}, and $M_2<M_1$. The diagonal dotted line corresponds to the $\Lambda_1=\Lambda_2$ case, in which $M_1=M_2$. The mass of the companion star is calculated through the relationship between $M_1$, $M_2$ and the chirp mass $\mathcal{M} = (M_1M_2)^{3/5}/(M_1+M_2)^{1/5}=1.188M_\odot$~\cite{ligo17}, i.e., $1.17 \leqslant M_2/M_\odot \leqslant 1.36$~\cite{ligo17,ligo18}. The upper and lower orange lines of the figure correspond to the 90\% and 50\% confidence limits, respectively, also obtained from the GW170817 event~\cite{ligo18}. From this figure, we also verify that the inclusion of DM in the system moves the bands in the direction of satisfying the LVC constraints. Notice that the system in which $k_F^{\rm DM}=0.03$~GeV is totally compatible with the 90\% region for all values chosen for $L_0$ in the range of Eq.~(\ref{sloperange}). 
These general features presented in Fig.~\ref{def} are also observed in Refs.~\cite{rmfdm3,eftdm1}, where other RMF-DM models are used (without SRC) including a chiral effective hadronic model. Therefore, our results point out to a particular pattern regarding RMF models with DM included, at least concerning the tidal deformability. However, it is important to mention that by increasing the amount of DM  can also enhance~$\Lambda$. This is the case of some models in which a dark matter halo~\cite{nelson,ellis,sagun} is generated. In Ref.~\cite{nelson}, for instance, it is verified an increase of $\Lambda$ for a total dark matter TOV mass~($M_{\mbox{\tiny DM}}$) exceeding $10^{-5}M_\odot$. In the analysis performed in Ref.~\cite{sagun}, bosonic self-interaction dark matter is coupled to a hadronic model through a two-fluid formalism (different from that used in this work). It is shown that for DM particle masses smaller than~$\sim 300$~MeV, $\Lambda$ increases with the DM fraction (here we fix the fermionic DM particle mass in $200$~GeV). Finally, in Ref.~\cite{ellis} the authors show that in the case of formation of dark matter halo, the tidal deformability increases by raising~$M_{\mbox{\tiny DM}}$. The opposite situation is verified in the case of a neutron star with a DM core. 

For the neutron stars described in the aforementioned works, a more sophisticated treatment of the contribution of the dark matter  is performed, namely, the DM Fermi momentum is not taken as constant along the star radius. In our work, we implement the simpler case of fixing $k_F^{\mbox{\tiny DM}}$ as also performed in Refs.~\cite{rmfdm3,eftdm1}, for instance. However, this latter treatment is completely appropriate for the purposes of the present study, namely, the investigation of tidal deformabilities and its relation with the moment of inertia. 

We also performed an additional analysis by taking into account those \mbox{RMF-SRC-DM} parametrizations with a different range for the symmetry energy slope, namely, $40\mbox{ MeV}\leqslant L_0 \leqslant 60\mbox{ MeV}$, value often predicted by some hadronic models. We verified that these specific parametrizations are also compatible with the LIGO/Virgo predictions presented in Fig.~\ref{def}.

\begin{figure}[!htb] 
\centering
\includegraphics[scale=0.34]{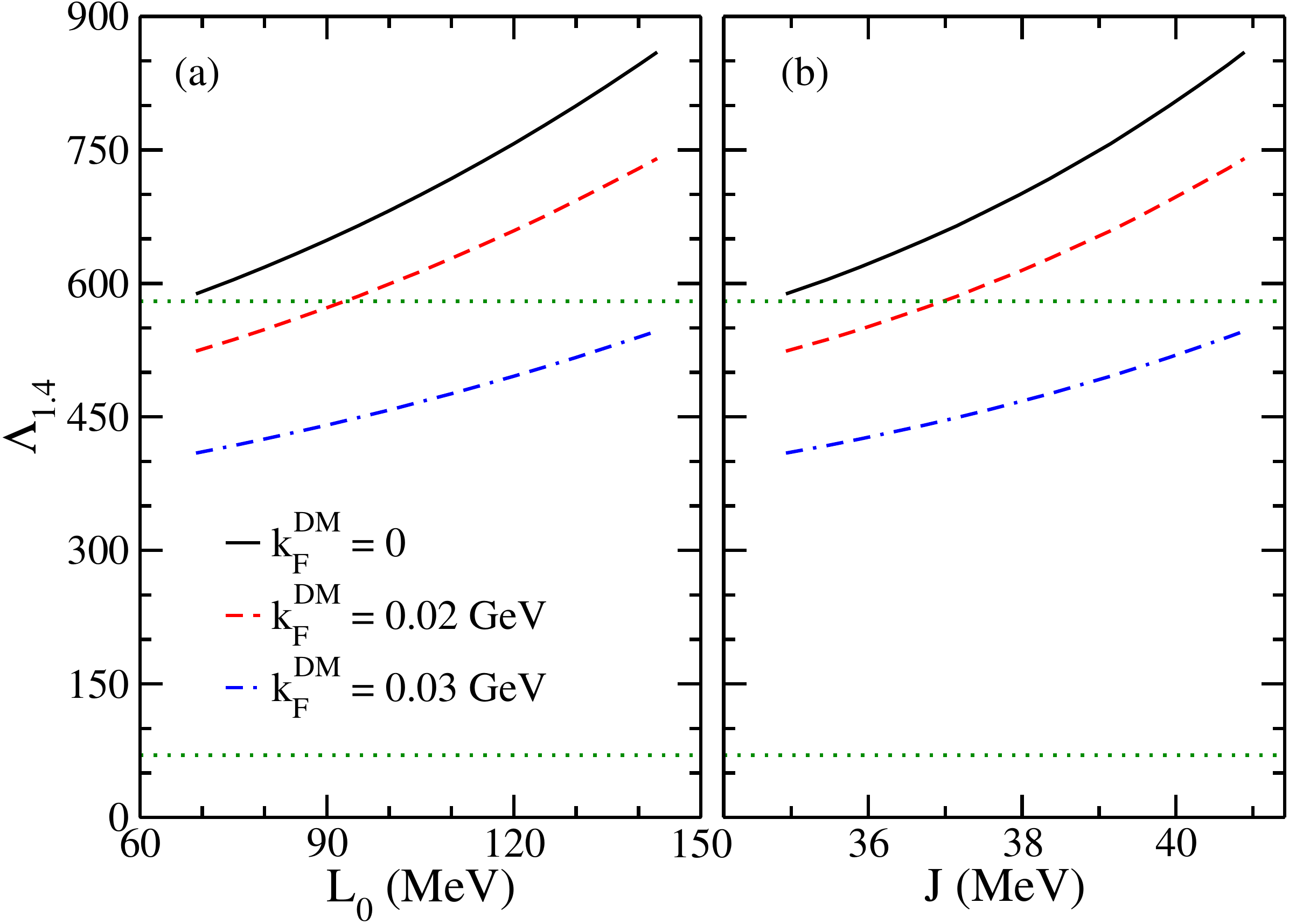}
\caption{$\Lambda_{1.4}$ as a function of (a) $L_0$ and (b) $J$ for different values of $k_F^{\rm DM}$. The range of $L_0$ is defined in Eq.~(\ref{sloperange}). Green dotted horizontal lines: boundaries of $\Lambda_{1.4}=190^{+390}_{-120}$~\cite{ligo18}.} 
\label{def14}
\end{figure}

In order to identify, in another perspective, the effect on $\Lambda_{1.4}$ of the DM content of the parametrizations generated from Eq.~(\ref{sloperange}), we show in Fig.~\ref{def14} how $\Lambda_{1.4}$ correlates with the isovector quantities $L_0$ and $J$ by taking into account different values of the dark particle Fermi momentum.
In Fig.~\ref{def14}{\color{blue}a} we see that $\Lambda_{1.4}$ decreases as $k_F^{\rm DM}$ increases, regardless the value of $L_0$.  The same occurs in Fig.~\ref{def14}{\color{blue}b}, now with respect to the symmetry energy at the saturation density. Notice that the dependence of $\Lambda_{1.4}$ on $L_0$ and $J$ reinforce the existence of a linear correlation between these two isovector quantities. Concerning $\Lambda_{1.4}\times L_0$, we remark that this pattern is also observed in a study performed in Ref.~\cite{lucas} in which hadronic model with SRC but without DM was analyzed. However, notice that the inclusion of DM content reduces the increasing of $\Lambda_{1.4}$ as function of $L_0$, since we have $\Delta\Lambda_{1.4}\equiv \Lambda_{1.4}(143)-\Lambda_{1.4}(69)$ given by $272$, $216$ and $138$, respectively, for $k_F^{\rm DM}=0$, $0.02$~GeV and $0.03$~GeV. We also remark that $\Lambda_{1.4}$ as an increasing function of $J$ in our analysis is totally different from the correlation exhibited in Ref.~\cite{lucas}. In that study, the authors considered independent variations of $J$ and $L_0$ and observed a decrease of $\Lambda_{1.4}$ with increase of $J$. Here, the opposite behavior is verified due to the linear correlation presented between $L_0$ and $J$. This relationship emerges since we are forcing a crossing point in the density dependence of the symmetry energy density. As aforementioned, we impose a value of $25.68$~MeV for the symmetry energy at $\rho\simeq 0.1$~fm$^{-3}$. We address the reader to Ref.~\cite{bianca} for a detailed study concerning crossing points and linear correlations of nuclear matter bulk parameters. Moreover, we emphasize that the relationship between $J$, $L$ and tidal deformabilities has been subject of investigation in many other works, such as those pointed out in Refs.~\cite{baoanli,baoanli2,baoanli3,malik19,cpc,fanji,sinha,ptep,liu,angli}.

\begin{figure}[!htb] 
\centering
\includegraphics[scale=0.34]{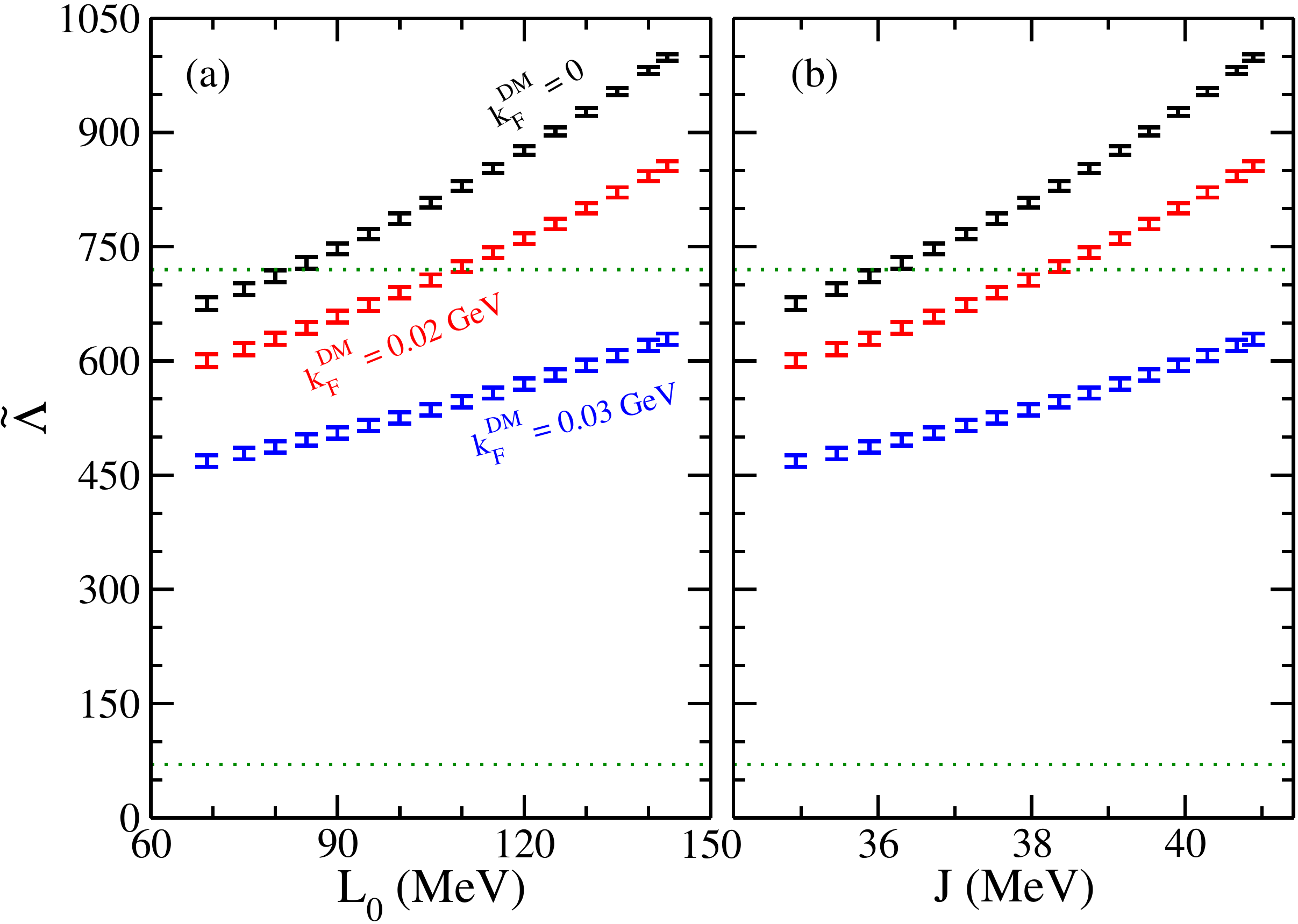}
\caption{$\tilde{\Lambda}$ as a function of (a) $L_0$ and (b) $J$ for different values of $k_F^{\rm DM}$. The range of $L_0$ is defined in Eq.~(\ref{sloperange}). Dashed lines: range of $\tilde{\Lambda}=300^{+420}_{-230}$ determined by LVC~\cite{ligo19}.} 
\label{deftilde}
\end{figure}

A quantity directly related to the tidal deformabilities of a binary NS system is the coefficient $\tilde{\Lambda}$ defined as
\begin{eqnarray}
    {\tilde{\Lambda}} = {16\over{13}}{{(M_{1}+12M_{2})M_{1}^{4}\Lambda_{1} + (M_{2}+12M_{1}) M_{2}^{4}\Lambda_{2}} \over {(M_{1}+M_{2})^{5}}},\qquad
\label{tilde}
\end{eqnarray}
where $\Lambda_{1}$ and $\Lambda_{2}$ are the dimensionless tidal deformabilities of each star.  In the inspiral final phase of a binary colliding NS system, periodic gravitational waves are emitted. The phase of these waves can be expressed in terms of a post-Newtonian expansion yielding a term proportional to $\tilde{\Lambda}$ at the lowest order~\cite{FH}. This result is used in order to investigate the response of the stellar system to the tidal field, being extracted directly from the observed waveform. In Fig.~\ref{deftilde} we show the plots $\tilde{\Lambda}\times L_0$ and $\tilde{\Lambda}\times J$ generated through the \mbox{RMF-SRC} model with different DM contents.
In this figure, $\tilde{\Lambda}$ is calculated as a function of the mass of one of the stars of the binary system, i.e., $\tilde{\Lambda}=\tilde{\Lambda}(M_1)$, or $\tilde{\Lambda}=\tilde{\Lambda}(M_2)$. As $M_1$, or $M_2$, is defined into a particular range according to the GW170817 event, each parametrization presenting a specific value of $L_0$, or $J$, produce a range for $\tilde{\Lambda}$. We compare the results obtained for the model with $k_F^{\rm DM}=0$, $0.02$~GeV, and $0.03$~GeV with the constraint $\tilde{\Lambda}=300^{+420}_{-230}$ provided by LVC~\cite{ligo19}. Once again, we notice that the inclusion of DM in the system favors the observational data from the GW170817 event. The decreasing of $\tilde{\Lambda}$ as a function of $k_F^{\rm DM}$ is also observed. Furthermore, as well as the behavior between $\Lambda_{1.4}$ and $L_0$ depicted in Fig.~\ref{def14}, there is also a strong correlation between $\tilde{\Lambda}$ and $L_0$. The same is true for the relationship between $\tilde{\Lambda}$ and $J$.

\begin{figure}[!htb]
\centering
\vspace{-1.2cm}
\includegraphics[width=0.54\textwidth]{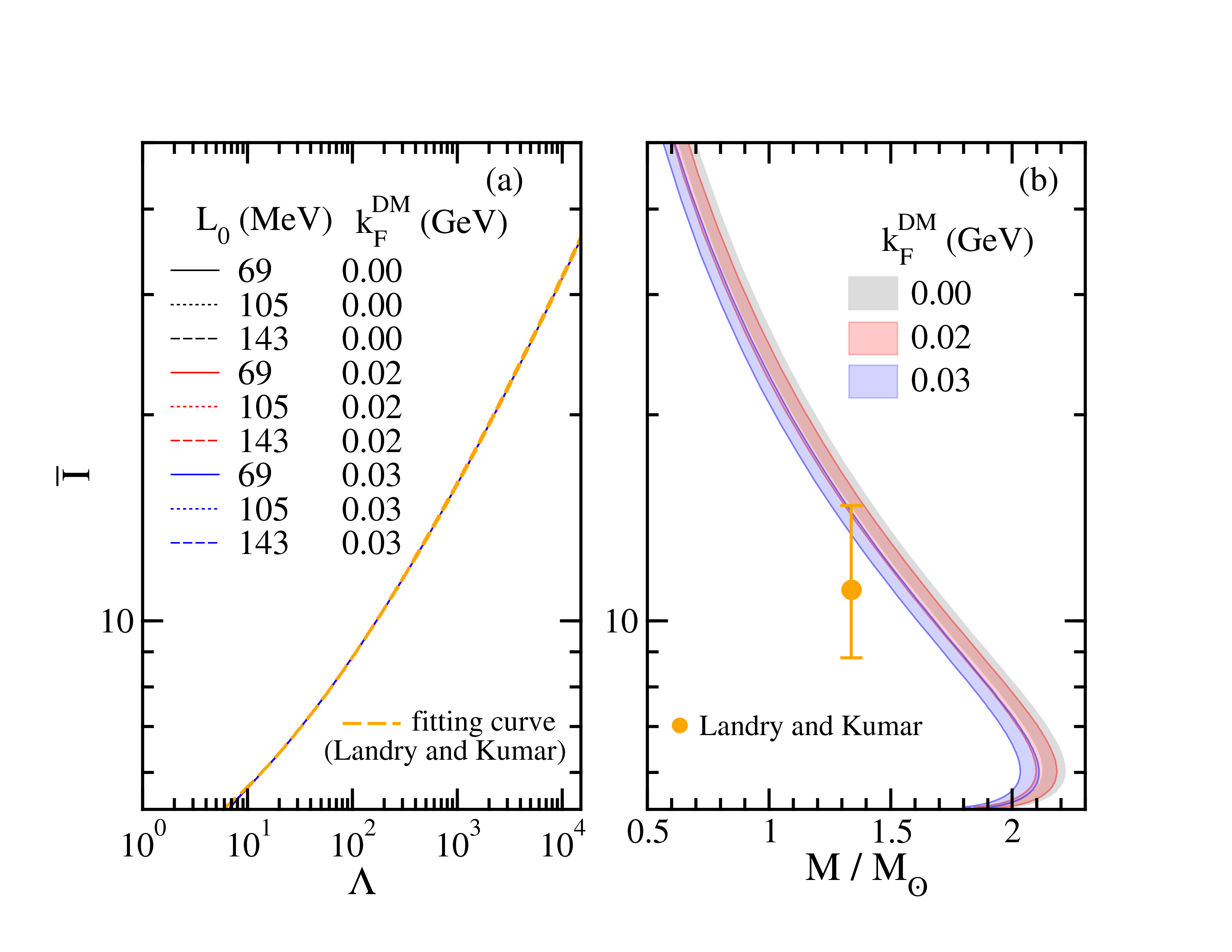}
\vspace{-1cm}
\caption{Dimensionless moment of inertia as a function of (a)~dimensionless tidal deformability, and (b)~the ratio $M/M_\odot$. Dashed curve: fitting curve obtained in Ref.~\cite{land18}. The circle with error bars represents an indirect prediction of $\bar{I}(M=1.338M_\odot)$ made in Ref.~\cite{land18} by considering the observational data of dimensionless tidal deformability (see text for more details).}
\label{inertia}
\end{figure}

As a last result, we show in Fig.~\ref{inertia} the dimensionless moment of inertia, $\bar{I}= I/M^3$, calculated from the \mbox{RMF-SRC} model for different values of $k_F^{\rm DM}$. This quantity is determined from the solution of the Hartle's slow rotation equation~\cite{land18,hartle,yagi13}. It is a differential equation for one of the metric decomposition functions~\cite{yagi13}, $\omega(r)$, coupled to the TOV equations. The moment of inertia is defined in terms of $\omega_R\equiv \omega(R)$ as $I=R^3(1-\omega_R)/2$. $\omega_R$ is the frame-dragging function evaluated at the star surface~\cite{land18}. 
The authors of Refs.~\cite{science,yagi13} showed that the relation between $\bar{I}$ and $\Lambda$ is independent of the neutron/quark star structure in the case of slowly-rotating stars. In Ref.~\cite{land18} the same result was obtained for a set of~$53$ Skyrme and RMF parametrizations. In Fig.~\ref{inertia}{\color{blue}a} it is clear that the parametrizations generated by the variation in Eq.~(\ref{sloperange}) are indistinguishable regardless the value of $k_F^{\rm DM}$. Therefore, we can conclude that the universal relation between $\bar{I}$ and $\Lambda$, called \mbox{$I$-Love} relation, is preserved even with the inclusion of dark matter in the system. The dashed line in Fig.~\ref{inertia}{\color{blue}a} represents the fitting curve determined in 
Ref.~\cite{land18}. We see that the model with DM  studied here is compatible with this fitting.

The authors of Ref.~\cite{land18} also determined a range for $\bar{I}$ related to the \mbox{PSR J0737-3039} primary component pulsar, namely, $\bar{I}_\star\equiv\bar{I}(M_\star)=11.10^{+3.68}_{-2.28}$, with $M_\star=1.338M_\odot$. This range was determined by using Skyrme and RMF parametrizations. Initially, it was verified a relation between $\Lambda_\star$ ($\Lambda$ related to $M_\star$) and $\Lambda_{1.4}$ (\mbox{binary-Love} relation). Then, a fitting for the $\Lambda_\star\times \Lambda_{1.4}$ curve was used with the \mbox{$I$-Love} relation in order to determine $\bar{I}_\star$ as a function of~$\Lambda_\star$. Lastly, the observational range $\Lambda_{1.4}=190^{+390}_{-120}$ from LVC was used to establish the limits for $\Lambda_\star$, and consequently, the range $\bar{I}_\star=11.10^{+3.68}_{-2.28}$. In Fig.~\ref{inertia}{\color{blue}b} we verify that the increase of $k_F^{\rm DM}$ produces a decrease of $\bar{I}$. We also find that the system with $k_F^{\rm DM}=0.03$~GeV is completely inside the limits for the moment of inertia of pulsar \mbox{PSR J0737-3039A} predicted in Ref.~\cite{land18}. Furthermore, we verify that parametrizations generated by the \mbox{RMF-SRC} model, i.e., with no dark matter, are in agreement with the mass-radius diagrams obtained from chiral effective theory calculations performed in Refs.~\cite{hebeler,kruger,drischler}, for $R\lesssim 14$~km. Curiously, on the other hand, the compatibility is fully attained with inclusion of dark matter content, specifically for $k_F^{\rm DM}=0.03$~GeV. In summary, this specific content of dark matter, implemented in the RMF model with short-range correlations, is compatible with all constraints derived from the GW170817 event concerning tidal deformabilities and moment of inertia.

\section{Summary and concluding remarks} 
\label{summ}

In this work, we investigate the capability of a hadronic relativistic model, with short-range correlations and dark matter content included~\cite{dmnosso}, in reproducing the observational data provided by the LIGO and Virgo Collaboration regarding the binary neutron star system of the GW170817 event, i.e., the one in which gravitational waves emitted from neutron stars merger were detected. We use the lightest neutralino, interacting with nucleons through the exchange of the Higgs boson, as the dark particle. In Ref.~\cite{dmnosso} it was already show that this model also reproduces the recent observational data obtained by the NICER mission~\cite{nicer1,nicer2,nicer3,nicer4}.

We show that the dimensionless tidal deformability~$\Lambda$ decreases as the Fermi momentum of the dark particle increases. In particular, this feature favors the model in satisfying the constraints of $\Lambda_{1.4}=190^{+390}_{-120}$ and $\tilde{\Lambda}=300^{+420}_{-230}$. Furthermore, a clear correlation between $\Lambda_{1.4}$ and the symmetry energy slope, $L_0$, and between $\tilde{\Lambda}$ and $L_0$ is verified for different values of $k_F^{\rm DM}$. Specifically, we use the variation of $L_0=(106\pm37)$~MeV~\cite{piekaprex2}, compatible with the updated results provided by the \mbox{PREX-2} collaboration concerning neutron skin thickness measurements of $^{208}\rm Pb$~\cite{prex2}, and also overlapping with the range found from the analysis of charged pions spectra~\cite{pions}. We also show that the $\Lambda_1\times\Lambda_2$ curves are moved to the direction of the GW170817 observational data.

Finally, we also analyze that the \mbox{$I$-Love} relation, namely, the relationship between $\Lambda$ and dimensionless moment of inertia, $\bar{I}$, is preserved even with the inclusion of dark matter in the system. The constraint of $\bar{I}_\star\equiv\bar{I}(M_\star)=11.10^{+3.68}_{-2.28}$, with $M_\star=1.338M_\odot$, is also satisfied for the system with $k_F^{\rm DM}=0.03$~GeV.

\section*{ACKNOWLEDGMENTS}
This work is a part of the project INCT-FNA proc. No. 464898/2014-5. It is also supported by Conselho Nacional de Desenvolvimento Cient\'ifico e Tecnol\'ogico (CNPq) under Grants No. 312410/2020-4 (O.L.), No. 308528/2021-2 (M.D.), and 308486/2015-3 (T.F.). We also acknowledge Funda\c{c}\~ao de Amparo \`a Pesquisa do Estado de S\~ao Paulo (FAPESP) under Thematic Project 2017/05660-0 and Grant No. 2020/05238-9 (O.L., C.H.L, M.D.).


\begin{thebibliography}{99}

\bibitem{einstein1} A. Einstein, Sitzungsber. K. Preuss. Akad. Wiss. {\bf 1}, 688 (1916).

\bibitem{einstein2} A. Einstein, Sitzungsber. K. Preuss. Akad. Wiss. {\bf 1}, 154 (1918).

\bibitem{bholes1} B. P. Abbott {\it et al}. (LIGO Scientific Collaboration and Virgo Collaboration), Phys. Rev. Lett. {\bf 116}, 061102 (2016); 

\bibitem{ligo17} B. P. Abbott {\it et al.} (LIGO Scientific Collaboration and Virgo
Collaboration), Phys. Rev. Lett. {\bf 119}, 161101 (2017).

\bibitem{dmrev} A. Arbey, and F. Mahmoudi, Prog. in Part. Nucl. Phys. {\bf 119}, 103865 (2021); G. Bertonea, D. Hooperb, and J. Silk, Phys. Rep. {\bf 405}, 279 (2005).

\bibitem{zwicky} F. Zwicky, Helv. Phys. Acta {\bf 6}, 110 (1933).

\bibitem{oort} J. H. Oort, Bull. Astr. Inst. Netherlands {\bf 6}, 249 (1932).

\bibitem{lensing} L. V. Koopmans, T. Treu, Astrophys. J. {\bf 583}, 606 (2003).

\bibitem{rmfdm1} A. Das, T. Malik, A. C. Nayak, arxiv:2011.01318.

\bibitem{rmfdm2} G. Panotopoulos and I. Lopes, Phys. Rev. D {\bf 96}, 083004 (2017).

\bibitem{rmfdm3} A. Das, T. Malik, and A. C. Nayak, Phys. Rev. D {\bf 99}, 043016 (2019).

\bibitem{rmfdm4} S. A. Bhat, A. Paul, Eur. Phys. J. C {\bf 80}, 544 (2020).

\bibitem{rmfdm5} A. Quddus, G. Panotopoulos, B. Kumar, S. Ahmad and S. K. Patra, J. Phys. G {\bf 47}, 095202 (2020).

\bibitem{rmfdm6} H. C. Das, A. Kumar, B. Kumar, K. Biswal, T. Nakatsukasa, A. Li and S. K. Patra, Mon. Not. R. Astron. Soc. {\bf 495}, 4893 (2020).

\bibitem{rmfdm7} H. C. Das, A. Kumar, B. Kumar, S. K. Biswal and S. K. Patra, J. Cosmol. Astropart. Phys. {\bf 01}, 007 (2021).

\bibitem{rmfdm8} H. C. Das, A. Kumar, and S. K. Patra, Mon. Not. R. Astron. Soc. {\bf 507}, 4053 (2021).

\bibitem{rmfdm9} A. G. Abac, C. C. Bernido, J. P. H. Esguerra, arxiv:2104.04969.

\bibitem{rmfdm10} H. C. Das, A. Kumar, S. K. Biswal, and S. K. Patra, Phys. Rev. D {\bf 104}, 123006 (2021).

\bibitem{rmfdm11} H. C. Das, A. Kumar, and S. K. Patra, Phys. Rev. D {\bf 104}, 063028 (2021).

\bibitem{rmfdm12} A. Kumar, H. C. Das, and S. K. Patra, Mon. Not. R. Astron. Soc. {\bf 513}, 1820 (2022).

\bibitem{cand1} J. L. Feng, Annu. Rev. Astro. Astrophys. {\bf 48}, 495 (2010).

\bibitem{cand2} A. Kusenko and L. J. Rosenberg, arxiv:1310.08642.

\bibitem{dmnosso} O. Louren\c{c}o, T. Frederico, and M. Dutra, Phys. Rev. D {\bf 105} 023008 (2022).

\bibitem{nature} M. Duer, O. Hen, E. Piasetzky, {\it et al.}, Nature {\bf 560}, 617 (2018). 

\bibitem{hen2017} O. Hen, G. A. Miller, E. Piasetzky and L. B. Weinstein, Rev. Mod. Phys. {\bf 
89}, 045002 (2017).

\bibitem{Duer2019} M. Duer, O. Hen, E. Piasetzky, {\it et al.}, Phys. Lett. B {\bf 797}, 134792 
(2019).

\bibitem{rev3} B. A. Li, L. W. Chen, and C. M. Ko, Phys. Rep. {\bf 464}, 113 (2008).

\bibitem{cai} B. J. Cai and B. A. Li, Phys. Rev. C  {\bf 93}, 014619 (2016).

\bibitem{baoanlicai} B. J. Cai, and B. A. Li, arxiv:2203.12773.

\bibitem{orhen} O. Hen, {\it et al.}, Science {\bf 346}, 614 (2014).

\bibitem{subedi2008} R. Subedi, {\it et al.}, Science {\bf 320}, 1476 (2008).

\bibitem{cromartie} H. T. Cromartie, E. Fonseca, S. M. Ransom, P. Demorest {\it et al.}, Nature Astronomy {\bf 4}, 72 (2020).

\bibitem{nicer1} M. T. Wolff {\it et al}., Astrophys. J. Lett. {\bf 918}, L26 (2021).

\bibitem{nicer2} M. C. Miller {\it et al}., Astrophys. J. Lett. {\bf918}, L28 (2021).

\bibitem{nicer3} T. E. Riley {\it et al}., Astrophys. J. Lett. {\bf 918}, L27 (2021).

\bibitem{nicer4} G. Raaijmakers, S. K. Greif, K. Hebeler, T. Hinderer, S. Nissanke, A. Schwenk, T. E. Riley, A. L. Watts, J. M. Lattimer, and W. C. G. Ho, Astrophys. J. Lett. {\bf 918}, L29
(2021).

\bibitem{piekaprex2} B. T. Reed, F. J. Fattoyev, C. J. Horowitz, and J. Piekarewicz, Phys. Rev. Lett. {\bf 126}, 172503 (2021).

\bibitem{prex2} D. Adhikari {\it et al.}, Phys. Rev. Lett. {\bf 126}, 172502 (2021).

\bibitem{pions} J. Estee {\it et al.}, Phys. Rev. Lett. {\bf 126}, 162701 (2021).

\bibitem{cromartie-apj} E. Fonseca, H. T. Cromartie {\it et al.}, Astrophys. J. Lett. {\bf 915}, L12 (2021).

\bibitem{ligo18} B. P. Abbott {\it et al}. (LIGO Scientific Collaboration and Virgo Collaboration), Phys. Rev. Lett. {\bf 121}, 161101 (2018).

\bibitem{ligo19} B. P. Abbott {\it et al}. (LIGO Scientific Collaboration and Virgo
Collaboration), Phys. Rev. X {\bf 9}, 011001 (2019).

\bibitem{cline} J. M. Cline, P. Scott, K. Kainulainen, and C. Weniger, Phys. Rev. D {\bf 88}, 055025 (2013); J. M. Cline, K. Kainulainen, P. Scott, and C. Weniger, Phys. Rev. D {\bf 92}, 039906(E) (2015).

\bibitem{pandaxII} A. Tan {\it et al}. (PandaX-II Collaboration), Phys. Rev. Lett. {\bf 117}, 121303 (2016).

\bibitem{lux} D. S. Akerib {\it et al.} (LUX Collaboration), Phys. Rev. Lett. {\bf 118}, 021303 (2017).

\bibitem{darkside} L. Marini {\it et al.} (DarkSide Collaboration), Nuovo Cim. C {\bf 39}, 247 (2016).

\bibitem{lucas} L. A. Souza, M. Dutra, C. H. Lenzi and O. Louren\c{c}o, Phys. Rev. C {\bf 101}, 065202 (2020).

\bibitem{tov39} R. C. Tolman, Phys. Rev. {\bf 55}, 364 (1939).

\bibitem{tov39a} J. R. Oppenheimer and G. M. Volkoff, Phys. Rev. {\bf 55}, 374 (1939).

\bibitem{bps} G. Baym, C. Pethick, and P. Sutherland, Astrophys. J. {\bf 170}, 299 (1971).

\bibitem{poly0} B. Link, R. I. Epstein, and J. M. Lattimer, Phys. Rev. Lett. {\bf 83} (1999) 3362.

\bibitem{poly1} J. Carriere, C. Horowitz, and J. Piekarewicz, Astrophys. J. {\bf 593} (2003) 463.

\bibitem{poly2} J. Piekarewicz, and F. J. Fattoyev, Phys. Rev. C {\bf 99} (2019) 045802.

\bibitem{gogny2} C. Gonzalez-Boquera, M. Centelles, X. Vi\~nas, and L. M. Robledo, Phys. Lett. B 
{\bf 779}, 195 (2018).

\bibitem{cc2} J. Xu, L.-W. Chen, B.-A. Li, and H.-R. Ma, Astrophys. J. {\bf 697}, 1549 (2009).

\bibitem{gogny1} J. Decharg\'e and D. Gogny, Phys. Rev. C {\bf 21}, 1568 (1980).

\bibitem{kubis04} S. Kubis, Phys. Rev. C {\bf 70}, 065804 (2004).

\bibitem{gonzalez19} C. Gonzalez-Boquera, M. Centelles, X. Vi\~nas, and T. R. Routray, Phys. Rev. C {\bf 100}, 015806 (2019).

\bibitem{tanj10} T. Hinderer, B. D. Lackey, Ryan N. Lang and J. S. Read, Phys. Rev. D
{\bf 81}, 123016 (2010).

\bibitem{read} J. S. Read, L. Baiotti, J.D.E. Creighton, J.L. Friedman, B. Giacomazzo, 
K. Kyutoku, C. Markakis, L. Rezzolla, M. Shibata, and K. Taniguchi, Phys. Rev. D {\bf 88}, 044042 (2013).

\bibitem{pozzo} W. Del Pozzo, T. G. F. Li, M. Agathos, C. VanDenBroeck, and S. Vitale, Phys. Rev. Lett. {\bf 111}, 071101 (2013).

\bibitem{new} S. Postnikov, M. Prakash, and J. M. Lattimer, Phys. Rev. D {\bf 82}, 024016 (2010).

\bibitem{hind08} T. Hinderer, Astrophys. J. {\bf 677}, 1216 (2008).

\bibitem{damour} T. Damour and A. Nagar, Phys. Rev. D {\bf 81}, 084016 (2010).

\bibitem{tayl09} T. Binnington and E. Poisson, Phys. Rev. D {\bf 80}, 084018 (2009).

\bibitem{fsu2r-new} L. Tolos , M. Centelles, and A. Ramos, Publ. Astron. Soc. Austral. {\bf 34}, e065 (2017).

\bibitem{pieka2001} C. J. Horowitz and J. Piekarewicz, Phys. Rev. Lett. {\bf 86}, 5647 (2001).

\bibitem{drischler} C. Drischler, R. J. Furnstahl, J. A. Melendez, and D. R. Phillips, Phys. Rev. Lett. {\bf 125}, 202702 (2020).

\bibitem{baoanli} B. A. Li, B. J. Cai, W. J. Xie, N. B. Zhang, Universe {\bf 7}, 182 (2021).

\bibitem{bianca} B. M. Santos, M. Dutra, O. Louren\c{c}o, and A. Delfino, Phys. Rev. C {\bf 92}, 015210 (2015).

\bibitem{wei} Jin-Biao Wei, Jia-Jing Lu, G. F. Burgio, Zeng-Hua Li, H.-J. Schulze, Eur. Phys. J. A {\bf 56}, 63 (2020).

\bibitem{eftdm1} D. Sen, and A. Guha, Mon. Not. R. Astron. Soc. {\bf 504}, 3354 (2021).

\bibitem{nelson} A. E. Nelson, S. Reddya, and D. Zhou, J. Cosmol. Astropart. Phys. {\bf 07}, 012 (2019).

\bibitem{sagun} D. R. Karkevandi, S. Shakeri, V. Sagun, and O. Ivanytskyi, Phys. Rev. D {\bf 105}, 023001 (2022).

\bibitem{ellis} J. Ellis, G. H\"utsi, K. Kannike, L. Marzola, M. Raidal, and V. Vaskonen, Phys. Rev. D {\bf 97}, 123007 (2018).

\bibitem{baoanli2} B. A. Li and M. Magno, Phys. Rev. C {\bf 102}, 045807 (2020).

\bibitem{baoanli3} P. G. Krastev and B. A. Li, J. Phys. G {\bf 46}, 074001 (2019).

\bibitem{malik19} T. Malik, B. K. Agrawal, J. N. De, S. K. Samaddar, C. Provid\^encia, C. Mondal, and T. K. Jha, Phys. Rev. C {\bf 99}, 052801(R) (2019).

\bibitem{cpc} Nai-Bo Zhang, Bin Qi, and Shou-Yu Wang, Chin. Phys. C {\bf 44}, 064103 (2022).

\bibitem{fanji} Fan Ji, Jinniu Hu, Shishao Bao, and Hong Shen, Phys. Rev. C {\bf 100}, 045801, (2019).

\bibitem{sinha} V. B. Thapa and M. Sinha, Phys. Rev. C {\bf 105}, 015802 (2022).

\bibitem{ptep} J. Hu, S. Bao, Y. Zhang, K. Nakazato, K. Sumiyoshi, and H. Shen, Prog. Theor. Exp. Phys. 043D01 (2020).

\bibitem{liu} H. Liu, J. Xu, and P. C. Chu, Phys. Rev. D {\bf 105}, 043015 (2022).

\bibitem{angli} Z. Miao, A. Li, Z. Zhu, and S. Han, Astrophys. J. {\bf 904}, 103 (2020).

\bibitem{FH} E. E. Flanagan and T. Hinderer, Phys. Rev. D {\bf 77}, 021502(R) (2008).

\bibitem{land18} P. Landry and B. Kumar, Astrophys. J. Lett. {\bf 868}, L22 (2018).

\bibitem{hartle} J. B. Hartle, Astrophys. J. {\bf 150}, 1005 (1967).

\bibitem{yagi13} K. Yagi, and N. Yunes, Phys. Rev. D {\bf 88}, 023009 (2013).

\bibitem{science} K. Yagi, and N. Yunes, Science {\bf 341}, 365 (2013).

\bibitem{hebeler} K. Hebeler, J. Lattimer, C. Pethick, and A. Schwenk, Astrophys. J. {\bf 773}, 11 (2013).

\bibitem{kruger} T. Kr\"uger, I. Tews, K. Hebeler, and A. Schwenk, Phys. Rev. C {\bf 88}, 025802 (2013).

\bibitem{drischler} C. Drischler, A. Carbone, K. Hebeler, and A. Schwenk, Phys. Rev. C {\bf 94}, 054307 (2016).


\end{thebibliography}
\end{document}